\documentclass{emulateapj}
 \usepackage{apjfonts}

\newcommand{\etal}{et al.}
\def\gsim{\lower 2pt \hbox{$\, \buildrel {\scriptstyle >}\over
{\scriptstyle \sim}\,$}}
\def\lsim{\lower 2pt \hbox{$\, \buildrel {\scriptstyle <}\over
{\scriptstyle \sim}\,$}}

\def\lya{Ly$\alpha$}

\def\ciii{C~{\scriptsize III}}
\def\civ{C~{\scriptsize IV}}

\def\nv{N~{\scriptsize V}}

\def\ovii{O~{\scriptsize VII}}
\def\ovi{O~{\scriptsize VI}}

\def\neviii{Ne~{\scriptsize VIII}}

\def\HI{H~{\scriptsize I}}
\def\hi{H~{\scriptsize I}}

\def\siiv{Si~{\scriptsize IV}}
\def\siiii{Si~{\scriptsize III}}

\RequirePackage[T1]{fontenc} 

\shortauthors{Yao \etal}
\shorttitle{Multiple Absorption Line Spectroscopy of IGM} 

\slugcomment{\em Accepted for publication in the  Astrophysical Journal}

\begin{document}

\title{Multiple Absorption-Line Spectroscopy of the \\
Intergalactic Medium. I. Model}
\author{Yangsen Yao\altaffilmark{1},
	J. Michael Shull\altaffilmark{1}, 
	Charles W. Danforth\altaffilmark{1},
	Brian A. Keeney\altaffilmark{1}, and
	John T. Stocke\altaffilmark{1}}
\altaffiltext{1}{Center for Astrophysics and Space Astronomy,
Department of Astrophysical and Planetary Sciences,
University of Colorado, 389 UCB, Boulder, CO 80309; yaoys@colorado.edu} 

\begin{abstract}
We present a physically-based absorption-line model for the 
spectroscopic study of the intergalactic medium (IGM). This model adopts 
results from \emph{Cloudy} simulations and theoretical calculations by 
\citet{gnat07} to examine the resulting observational signatures of the 
absorbing gas with the following ionization scenarios: 
collisional ionization equilibrium (CIE), photoionization equilibrium, 
hybrid (photo- plus collisional ionization), and non-equilibrium cooling. 
As a demonstration, we apply this model to new observations made with 
the {\it Cosmic Origins Spectrograph} aboard the {\it Hubble Space Telescope} 
of the IGM absorbers at $z\sim0.1877$ along the 1ES~1553+113 sight line. 
We identify Ly$\alpha$, \ciii, \ovi, and \nv\ absorption lines with two distinct 
velocity components (blue at $z_b = 0.18757$; red at $z_r = 0.18772$) 
separated by $\Delta (cz)/(1+z) \approx 38~{\rm km~s}^{-1}$.   Joint analyses 
of these lines indicate that none of the examined ionization scenarios can 
be applied with confidence to the blue velocity component, although photoionization 
seems to play a dominant role. For the red component, CIE can be ruled out, but 
pure photoionization and hybrid scenarios (with $T < 1.3 \times10^5$ K) are more
acceptable. The constrained ranges of hydrogen density and metallicity of the 
absorbing gas are $n_{\rm H} = (1.9-2.3)\times10^{-5}~{\rm cm^{-3}}$ and 
$Z = (0.43-0.67)Z_\odot$.  These constraints indicate \ovi\ and \hi\ ionization 
fractions, $f_{\rm OVI} = 0.10-0.15$ and $f_{\rm HI} = (3.2-5.1)\times10^{-5}$,
with total hydrogen column density $N_H = (0.7-1.2)\times10^{18}~\rm{cm^{-2}}$. 
This demonstration shows that joint analysis of multiple absorption lines can constrain 
the ionization state of an absorber, and results used to estimate the baryonic matter 
contained in the absorber. 

\end{abstract}


\section{Introduction }
\label{sec:intro}

The intergalactic medium (IGM) is believed to be the main reservoir of baryons 
throughout the history of the universe (e.g., \citealt{shu09}). 
In the early universe ($z>3$) the IGM was dominated by photoionized gas traced 
by Ly$\alpha$ forest lines (e.g., \citealt{rau97}).   At $z\lsim2$, the IGM is expected 
to be a mixture of photoionized and collisionally ionized gas produced by radiation
of active galactic nuclei (AGN) and gravitational shock heating during the formation 
of galaxies, groups, and clusters (e.g., \citealt{birn03, ker09}). Cosmological 
hydrodynamic simulations of large-scale structure formation predict that, at the 
current epoch, $\gsim40\%$ of the baryons may exist in the form of a warm-hot 
intergalactic medium (WHIM) at temperatures $T\sim10^5-10^7$~K in a 
``Cosmic Web'' distributed between the galaxies (e.g., \citealt{cen06b, opp08}).

Conducting a complete baryon inventory for the current universe is one of the major 
tasks of modern cosmology.  The predicted physical conditions in the hottest parts of 
the WHIM may require detection using X-ray absorption lines (e.g., \citealt{dave01, 
cen06}).  However, owing to limited sensitivity and spectral resolution, X-ray observations 
have so far yielded only a few controversial detections 
(e.g., \citealt{fang02, fang07, nic05, kaa06, buo09}).   Current X-ray observatories 
such as {\sl Chandra} and {\sl XMM-Newton} have spectral resolution 
($\lambda/\Delta \lambda \sim 400$) and detection limits ($\log N_{\rm OVII} \gsim 15.5$; 
\citealt{yao09, yao10}).   Instruments operating in the far- and near-ultraviolet (UV) 
wavelength bands have superior spectral resolution, $\lambda/\Delta \lambda \sim 
20,000-43,000$.   With the larger intrinsic line strengths (\ovi\  vs.\ \ovii) and the
higher spectral resolution and sensitivity, the UV spectrographs aboard the 
{\it Hubble Space Telescope} ({\sl HST}) -- the Space Telescope Imaging Spectrograph 
(STIS) and the Cosmic Origins Spectrograph (COS) -- are sensitive to lower column
densities, $\log N_{\rm OVI}  \gsim 12.7$.   These instruments can probe the WHIM up to
redshifts $z  \lsim  0.7$ (e.g., \citealt{dan05, dan08, sav05, tri08, tho08}).   The major 
advances in low-redshift IGM studies in the next decade are expected to come from archival 
data from STIS and  the {\it Far Ultraviolet Spectroscopic Explorer} ({\sl FUSE}) together with 
new observations from COS.  Indeed, the multi-phase nature of the IGM has been revealed
 through detected absorption lines of ions at various ionization states (e.g., 
\hi, \ciii, \civ, \siiii, \siiv,  \nv, \ovi, and \neviii) by {\sl FUSE}, STIS, and COS.
In order to proceed, we need to know how to convert measurement of these 
absorption lines to amounts of baryonic matter, which depends on ionization 
fractions, metallicity, and ionization phases of the IGM (\citealt{pen04, tri08, dan08}, 
hereafter DS08).  None of these properties is well measured for many absorbers, 
and in many cases the data do not permit such determinations.

Spectroscopy of multiple absorption lines provides a powerful diagnostic of IGM properties. 
These diagnostics have been applied to many systems by measuring ionic column densities 
and doppler velocities, examining the significance of  photo-, collisional, and non-equilibrium 
ionization, and assessing the physical conditions in the absorbers 
(e.g., \citealt{ric04, sav05, tri08, nar10}).  However, because these applications follow a 
multistep procedure, the derived physical parameters could depend in complex ways on the
initially measured quantities (e.g., temperature and metallicity of an intervening 
gas could co-vary as different functions of \ovi\ column density),
which makes estimations of their uncertainty hard to implement.  We present an 
absorption-line model in which the
physical properties of the absorber have been used as the primary fitting
parameters. Spectral fitting of multiple absorption lines 
with this model will not only
yield constraints to these parameters but automatically take care
of covariance of all parameters. As a case study, we apply this model to 
an interesting and well-defined absorption system at $z\sim0.1877$
along the sight line of 1ES~1553+113 observed with COS.

The paper is organized as follows.
We describe our COS observations and data reduction procedure in 
\S~\ref{sec:obs}, and introduce our spectral model in 
\S~\ref{sec:models}. We apply our model to the observations
and discuss our results and their implications in \S~\ref{sec:dis}.

\section{Observations and data reduction}
\label{sec:obs}

One of the Guaranteed Time Observation (GTO) targets, 
the BL Lacertae Object 1ES~1553+113 was observed on 2009 September 22 with 
COS, which was installed on the {\sl HST} during Hubble Servicing Mission 4.
The observations were taken with the same FP\_POS position
(FP\_POS=3) but with different central wavelengths, shifting along the
sequence 1291 \AA, 1300 \AA, 1309 \AA, and 1318 \AA\ for the G130M grating
(covering 1134 \AA\ to 1465 \AA), and 1589 \AA, 1600 \AA, 1611 \AA, and
1623 \AA\ for the G160M grating (covering 1383 \AA\ to 1797 \AA). 
The total exposure times are 3109.6~s with the G130M and
3802.8~s with the G160M.

The observations were calibrated with the pipeline CALCOS (version 2.11f). 
Flat-fielding, alignment, and coaddition of the processed exposures
were carried out using IDL routines developed by the COS GTO team
specifically for COS FUV data\footnote{See {\tt
http://casa.colorado.edu/$\sim$danforth/costools.html} for our
coaddition and flat-fielding algorithm and additional discussion.}. 
Briefly, true flat-fielding of COS observations has yet to be developed,
but each exposure was corrected for narrow, $\sim15\%$-opaque shadows
from repellor grid wires. 
The one-dimensional map of grid-wire opacity was shifted from detector
coordinates to wavelength space and divided from the reduced
one-dimensional flux. Within the same grating (G130M or G160M), 
these spectra were then cross-correlated and 
combined to form an exposure-weighted coadded spectrum
(See \citealt{dan10b} for a detailed description
of observations and data reduction).

\begin{center}
\begin{figure*}
  \plotone{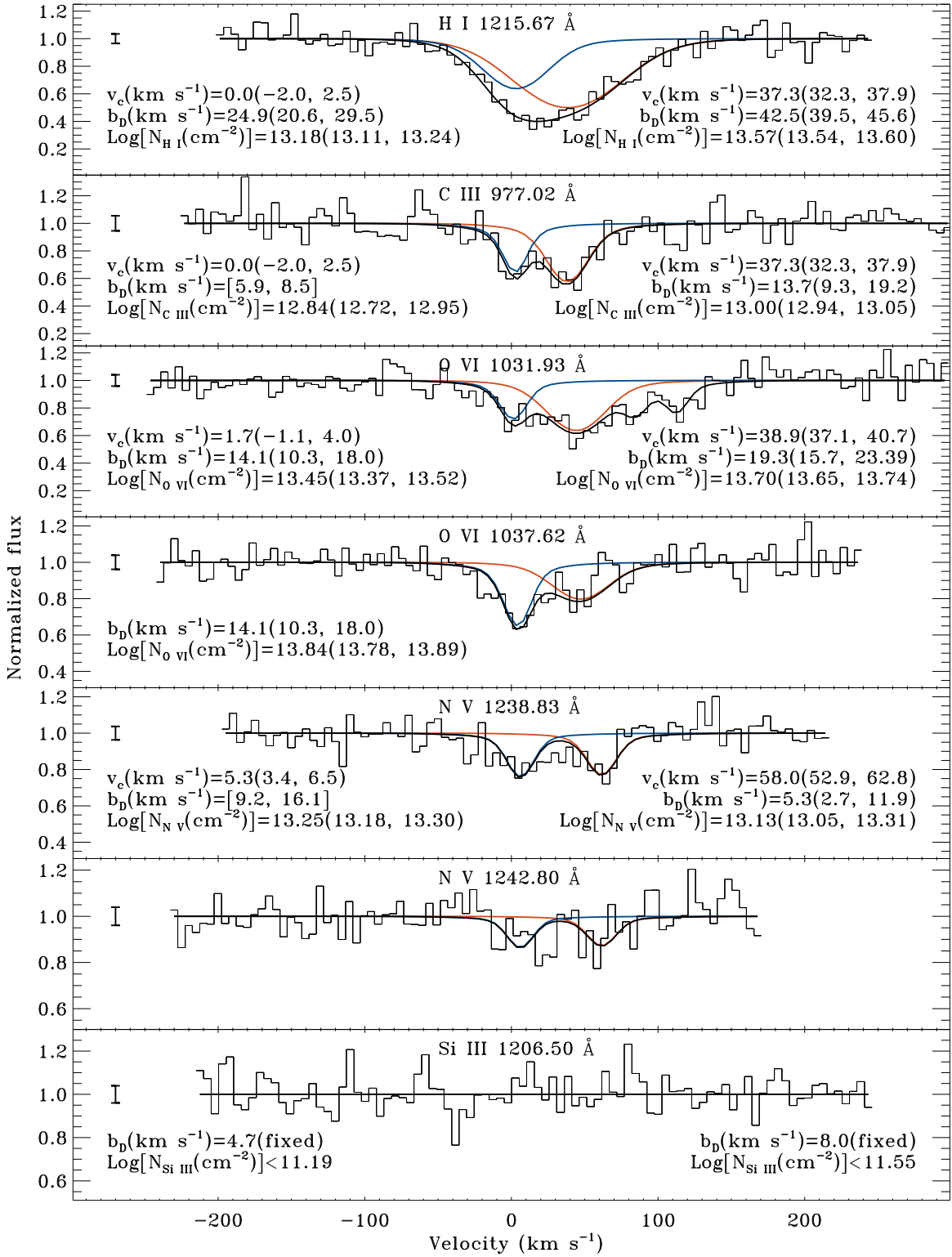}
  \caption{Absorption lines at $z\sim0.1877$ detected in the COS 
    spectrum of 1ES~1553+113. The vertical error bar shown at 
    $-270~{\rm km~s^{-1}}$ in each panel indicates the averaged flux error
    in the plotted wavelength range.
    Thick black lines show the continua and 
    total absorption, and blue and red lines mark fits to the two
    velocity components, at $z_b = 0.18757$ and $z_r = 0.18772$. 
    The labeled line shifts ($v_{\rm c}$),  doppler velocities ($b_{\rm D}$), 
    and column densities ($N_{\rm i}$) are measured relative to $z=0.18757$.
    Numbers in brackets ``[]'' are calculated values with respect to other
    ions. Doublet lines are jointly measured when applicable. 
    Data are binned by a factor of 2 with $\sim3$ bins in
    each resolution element. See text for details.
  }
  \label{fig:line_velocity}
\end{figure*}
\end{center}

In the COS spectra of 1ES~1553+113, three absorption systems were
identified near redshift $z=0.187$ \citep{dan10b}. Among these systems, 
the one at $z\sim0.1877$, as presented in Figure~\ref{fig:line_velocity},  
has the most significant detections of absorption lines
of Ly$\alpha$, \ciii, and the \ovi\ and \nv\ doublets. 
We therefore use this system to test our absorption-line model.

\section{Models}
\label{sec:models}

We considered intergalactic absorbers in the following ionization 
model conditions:
(1) collisional ionization equilibrium (CIE; Case 1);
(2) photoionization equilibrium (PE; Case 2);
(3) medium with a constant electron temperature illuminated by the UV
photoionizing background (hybrid model; Case 3); and
(4) post-shock gas undergoing non-equilibrium isobaric (Case 4)
or isochoric (Case 5) cooling. 
These five cases represent the most explored ionization
scenarios in ultraviolet spectroscopic studies of the IGM. 

For Cases 1, 4, and 5, we adopted the results of
\citet{gnat07}, in which they calculated 
the ionic abundances for gas in CIE state and 
for a shock-heated gas cooling from $\log T = 6.7$ either
at constant density (isochorically) or at constant pressure 
(isobarically). 

For Case 3, we ran {\sl Cloudy} (version 08.00;
\citealt{fer98}) simulations with parameters of temperature ($T$) and
hydrogen number density ($n_{\rm H}$). A 100-kpc thick slab 
(e.g., \citealt{dave01, pen04}) was used to
approximate an intergalactic absorbing cloud, 
which is illuminated by the intergalactic ionization field of
\citet{haa01} at redshifts $z=0-1$ 
\footnote{We in fact used the command ``HM05'' in Cloudy simulations to
  specify both 
  shape and flux of the ionization field; the normalization of the field
  evolves from $1.6\times10^{-6}~{\rm erg~s^{-1}~cm^{-2}}$ at $z=0$ to 
  $1.3\times10^{-5}~{\rm erg~s^{-1}~cm^{-2}}$ at $z=1$.}
with a step size of 0.1.
In our spectral fitting described below, the estimated ionization
field at the closest redshift bin to that of the observed absorber was 
utilized.  The grids of $\log T$ and $\log [n_{\rm H}({\rm cm^{-3}})]$ were
4 to 7.5 and $-6$ to $-2$, respectively, with a step size of 0.1 for each grid.
The collisional ionization effects were included by setting a constant
electron temperature in the simulations.
The relative ionic column densities were initially calculated with
the metallicity $\log (Z/Z_\odot) = -1$, and the results for metal
(heavier than helium) ions were then scaled to different metallicities. 
We took the results with $\log T =4$ for a pure photoionization model (Case 2).

\begin{deluxetable}{ll}
\tablewidth{0pt}
\tablecaption{Ionization Scenarios \label{tab:mod}}
\tablehead{\multicolumn{1}{c}{Scenarios} & Free Parameters}
\startdata
Case 1 : collisional ionization equilibrium (CIE)  &
$T$, $Z$, $N_{\rm H}$ \\
Case 2 : pure photoionization equilibrium (PE) &
$n_{\rm H}$, $Z$, $N_{\rm H}$ \\
Case 3 : photo- plus collisional ionization (hybrid) &
$n_{\rm H}$, $T$, $Z$, $N_{\rm H}$ \\
Case 4 : non-equilibrium isobaric cooling &
$T$, $Z$, $N_{\rm H}$ \\
Case 5 : non-equilibrium isochoric cooling &
$T$, $Z$, $N_{\rm H}$ 
\enddata
\tablecomments{Cases 1, 4, and 5 are based on results by \citet{gnat07},
  and Cases 2 and 3 are based on Cloudy simulations. 
  $n_{\rm H}$, $T$, $Z$, and $N_{\rm H}$ are the number density,
  temperature, metallicity, and total column density of the absorbing gas,
  respectively. }
\end{deluxetable}

\begin{figure}
  \plotone{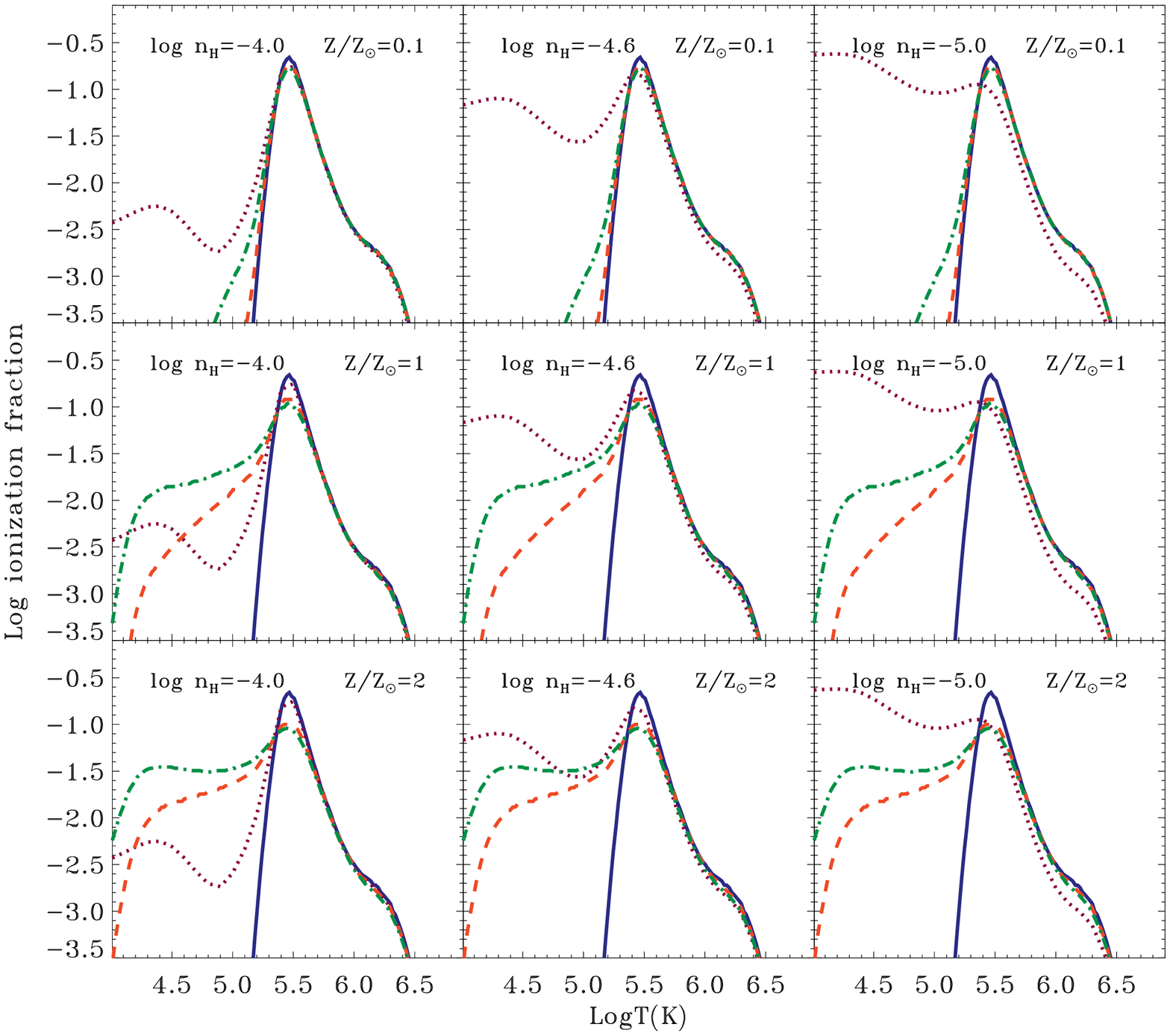}
  \caption{Ionization fraction of \ovi\ under different
  physical conditions. The solid-blue, dashed-red, dash-dotted-green, and
  dark-red-dotted curves indicate the CIE, Isobaric, Isochoric, and Hybrid
  models, respectively. For the hybrid model, plots show the 
  fraction in gas illuminated by the photoionization field at $z = 0$. 
}
  \label{fig:OVI_frac}
\end{figure}

\begin{figure}
  \plotone{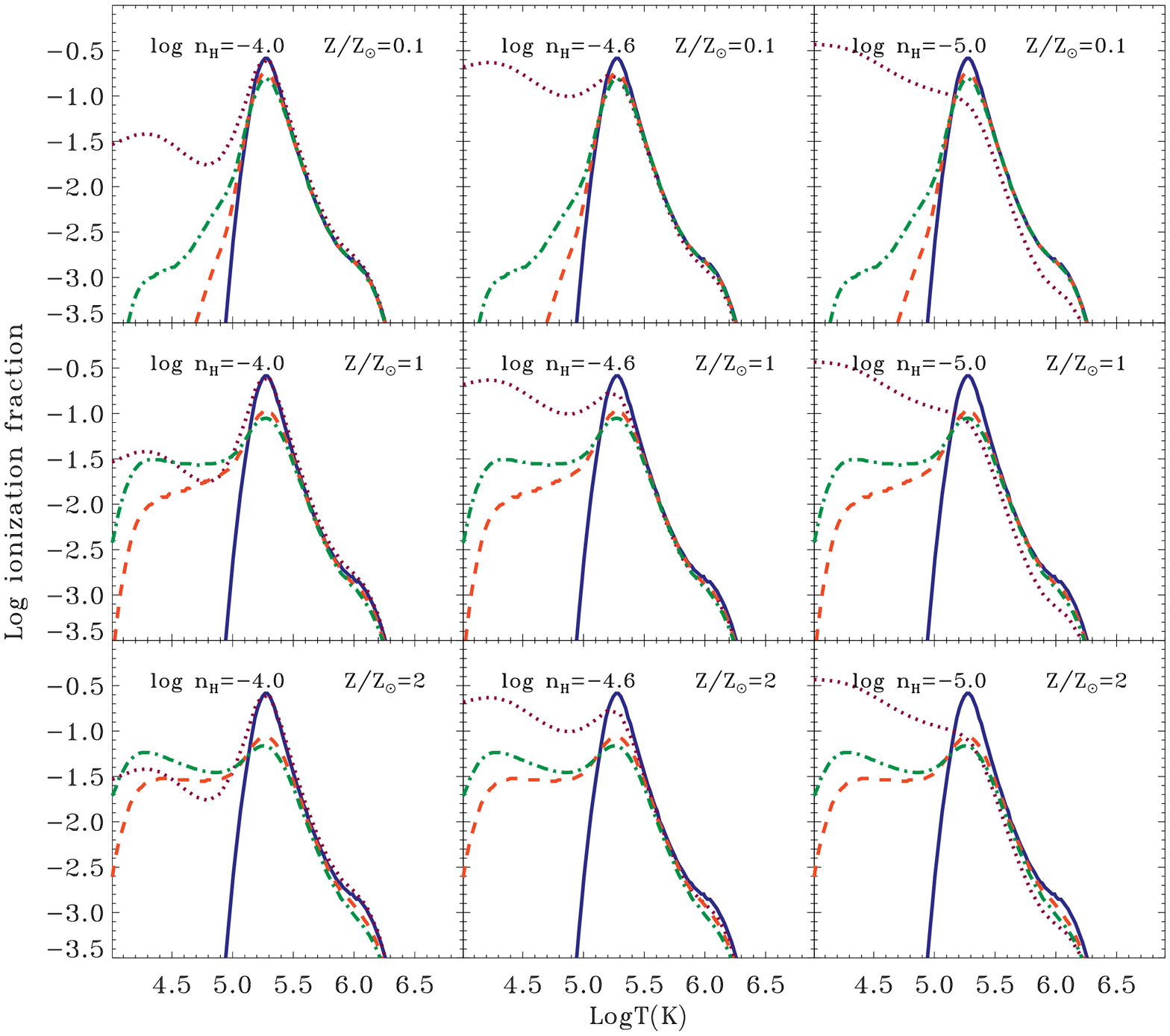}
  \caption{Same as Figure~\ref{fig:OVI_frac} but for ion \nv.
}
  \label{fig:NV_frac}
\end{figure}

Table~\ref{tab:mod} summarizes these ionization scenarios, and
Figures~\ref{fig:OVI_frac} and \ref{fig:NV_frac} demonstrate the 
differences in \ovi\ and \nv\ ionization fractions among different models.
In any of these models, for a given set of physical parameters 
(e.g., $Z$, $T$, and/or $n_{\rm H}$), there will be a determined set of 
column density ratios among all ions. 
In turn, if the observed multiple absorption lines occur in the same 
gas undergoing the same ionization scenario, the column density 
ratios among metal ions will provide constraints on $T$ and/or $n_{\rm H}$,
and ratios between metal ions and \hi\ will yield metallicity diagnostics of 
the absorbing gas.


We developed a model that can be used to jointly analyze multiple absorption 
lines to realize these diagnostics. The model, which is revised from our X-ray 
absorption-line model {\sl absline} \citep{yao05}, is summarized as follows. 
An absorption-line profile can be described as \citep{ryb79}: 
\begin{equation}
     f(\lambda) = f_c (\lambda)   \exp [ - \tau(\lambda)]  \; ,
\end{equation}
where $f_c$ is continuum flux and $\tau$ is the absorption optical depth,
given by a Voigt function depending on wavelength ($\lambda$),
column density ($N_i$), doppler width ($b_D$), 
oscillator strength ($f_{lu}$; $l$ and $u$ are the lower and upper
excitation levels for the transition), and the natural damping factor ($\gamma$)
of the transition. For a set of input parameters ($N_{\rm HI}$, $Z$, $T$,
and/or $n_{\rm H}$) of any ionization case discussed above, the model queries 
the simulation and/or calculation results and interpolates the values lying between 
the simulated and calculated grids to obtain a unique value for $N_i$. These 
physical properties are fitting parameters linked together in a joint
analysis of multiple lines. 

We implement our spectral analysis within the software package 
{\sl Xspec}\footnote{http://heasarc.nasa.gov/docs/xanadu/xspec/}
\citep{arn96}, to utilize its $\chi^2$ minimization algorithm over 
multiple fitting variables, where
\begin{equation}
  \chi^2=\sum_i\left(\frac{f_i^o-f_i^p}{\sigma_i}\right)^2  \;  .  
\end{equation}
Here, for each wavelength bin ($i$),  $f_i^o$, $f_i^p$, and $\sigma_i$ are the 
observed  flux, the model-predicted flux, and the uncertainty of $f_i^o$.
For COS gratings G130M and G160M, we interpolate the line spread 
functions\footnote{http://www.stsci.edu/hst/cos/performance/spectral\_resolution/}
(LSFs) to the desired wavelength grid and then use the IDL scripts 
{\sl wrt\_ogip\_rmf}\footnote{http://hea-www.harvard.edu/PINTofALE/pro/util/}
and {\sl writepha}\footnote{http://astro.uni-tuebingen.de/software/idl/aitlib/fits/} 
to transform these LSFs and the coadded spectra obtained in \S~\ref{sec:obs}
to a ``fits'' format for input to {\it Xspec}.   In spectral fitting, the spectral
models are automatically convolved with these wavelength-dependent LSFs.

In the following, we apply this model to the absorption system at $z=0.187$ along the 
sight line to 1ES~1553+113. Because velocity components in this system can clearly be
resolved (Fig.~\ref{fig:line_velocity}), the spectral analyses conducted 
here are similar to traditionally
employed profile-fitting techniques, although our physical variables are the fitting 
parameters.  When several velocity components are
severely blended, arbitrarily resolving components through spectral fitting
could lead to unreliable results.

\section{Spectral Fit, Results, and Discussion}
\label{sec:dis}

Using our absorption-line model, we first fit the lines individually to 
obtain properties of the absorbing ions. The local continua are interpolated
from the best fits to spectral ranges $\pm150~{\rm km~s^{-1}}$ from line centroids.
Absorption lines of \ciii, \ovi, and \nv\ clearly show two velocity components
(blue and red) in the $z\sim0.1877$ absorber, but component 
separation is not obvious in the broad Ly$\alpha$, other than asymmetry in 
the red wing (Figure~\ref{fig:line_velocity}).   In both components, centroids of 
the metal lines are well aligned, except for the red \nv\ that deviates
$\sim 20~{\rm km~s^{-1}}$ from others. We therefore set the centroid velocities 
of the two \hi\ components equal to those of the corresponding \ciii\ absorption. 
Fitting these lines individually yields the velocity shift $v_c$, $b_D$, and $N_i$ 
of the absorbing ions. For each  velocity component, we link $b_D$, $N_i$ and 
the centroids of the \ovi\ and \nv\ doublets to get better constraints (except
the blue component of \ovi). The \ovi\ 1037.62 \AA\ line should be about half 
the strength of \ovi\ 1031.93 \AA, but it appears stronger in the blue component. 
This mismatch is likely caused by an unidentified weak intergalactic 
Ly$\alpha$ absorber at $z=0.01364$ that contaminates the blue component of
\ovi\ 1037.62 \AA.   We therefore only link the $b_D$ and velocity centroids
and vary $N_{\rm OVI}$ independently in the doublet.   We also exclude this line 
in our joint analysis of multiple absorption lines described below. 
The red wing of the 1031.93 \AA\ line is blended with Ly$\beta$ 
absorption at $z=0.1950$. We use a two-velocity model to fit the Ly$\beta$ profile 
and fix its contribution in our analysis (Fig.~\ref{fig:line_velocity}).
The $b_D$ of \ciii\ and \nv\ cannot be constrained in the blue component, owing 
to the narrowness of the \ciii\ line and the poor S/N around the \nv\ line. We set the 
$b_D$ ranges for \ciii\ and \nv\ to those of \hi\ and \ovi\ assuming thermal broadening. 
Because there is no visible \siiii\ 1206 \AA\ line in the spectrum (see \citealt {dan10b}),
we fix the line centroid velocities to those of \ciii\ and $b_D$ to the 
corresponding thermal values of \hi\ to obtain upper limits on $N_{\rm SiIII}$. The 
constrained values of $v_c$, $b_D$, and $N_i$ 
are labeled in Figure~\ref{fig:line_velocity}.

Next, we jointly analyze these absorption lines in each velocity component 
to examine the five ionization scenarios discussed in \S~\ref{sec:models}.
Line widths of IGM absorbers have been shown to possess significant contributions 
from turbulent motion or unresolved components; measured widths therefore provide 
upper limits on thermal broadening in the absorber (\citealt{tho08, tri08, dan10a}).
Attributing the constrained $b_D$ to purely thermal broadening, we derive
upper limits of $\log T\le 4.7$ and 5.1 for the absorbing gas in Cases 1 and 3$-$5 
(both are based on \hi\ measurements) for blue and red components respectively. 
We apply the upper limits in our joint analyses, and constrain the physical properties 
($Z$, $T$, $n_{\rm H}$) and total hydrogen column density $N_{\rm H}$ of the absorbing 
gas (Table~\ref{tab:results}).  Given the constrained values of $Z$, $T$, $n_{\rm H}$, 
the column density of any ion can be calculated.  The predicted column densities
$\log~N_{\rm SiIII} = 8.82-11.16$ for all acceptable scenarios, consistent with the 
upper limits derived from the data (Figure~\ref{fig:line_velocity}; \citealt{dan10b}).  

\begin{deluxetable*}{c|ccccc|ccccc}
\tablewidth{0pt}
\tablecaption{Diagnostic Results \label{tab:results}}
\tablehead{
 & \multicolumn{5}{c}{Blue component $z_b = 0.18757$} & 
   \multicolumn{5}{c}{Red component $z_r = 0.18772$} \\
 & $\delta \chi^2$ & $\log n{\rm{_H}}$ & $\log T$ &
   $\log(Z/Z_\odot)$ & $\log N{\rm{_{H}}}$ & 
   $\delta \chi^2$ & $\log n{\rm{_H}}$ & $\log T$ &
   $\log( Z/Z_\odot)$ & $\log N{\rm{_{H}}}$ \\
 &  & (${\rm cm^{-3}}$) & (K) &
    & (${\rm cm^{-2}}$) & 
    & (${\rm cm^{-3}}$) &  (K) &
    & (${\rm cm^{-2}}$)}
\startdata
Case 1 & $>$100 & NA    & [$<$4.7] & $\times$ & $\times$  & $>$100 & NA & [$<$5.1] & $\times$ & $\times$ \\
Case 2 & 21.5 & $-4.69_{-0.04}^{+0.04}$ & NA    & $0.05_{-0.08}^{+0.09}$ & $17.51_{-0.06}^{+0.05}$ & 0.8 & $-4.74^{+0.03}_{-0.03}$ & NA    & $-0.22^{+0.05}_{-0.05}$ & $17.89^{+0.03}_{-0.03}$\\
Case 3 & 16.6 & $-4.62_{-0.08}^{+0.05}$ & [$<$4.7] & $0.07_{-0.09}^{+0.07}$ & $17.57_{-0.08}^{+0.09}$ & 0.1 & $-4.67^{+0.03}_{-0.09}$ & [$<$5.1] & $-0.24^{+0.07}_{-0.12}$ & $17.96^{+0.11}_{-0.10}$\\
Case 4 & $>$100 & NA    & [$<$4.7] & $\times$ & $\times$  & 4.1 & NA & ($>$5.0), [$<$5.1] & $0.13_{-0.01}^{+0.14}$ & $18.58^{+0.03}_{-0.16}$\\
Case 5 & $>$100 & NA    & [$<$4.7] & $\times$ & $\times$  & 3.9 & NA & ($>$5.0), [$<$5.1] & $0.02_{-0.02}^{+0.06}$ & $18.59^{+0.02}_{-0.06}$

\enddata
\tablecomments{Cases 1-5 are for gas in pure CIE, pure photoionization, 
  photoionization plus collisional ionization, non-equilibrium isobaric 
  cooling, and non-equilibrium isochoric cooling  states, respectively.
  $\delta\chi^2$ is the difference of the best-fit $\chi^2$ of a 
  joint analysis from that of fitting lines individually.
  The ``NA'' symbol indicates that the parameter is not applicable to the
  model, and the ``$\times$'' symbol indicates that the model can be ruled
  out by the data. Limits within parentheses ``()'' are obtained in joint
  analysis, and limits within brackets ``[]'' are derived in fitting
  individual lines. Errors are reported at 1$\sigma$ uncertainty. See text
  for details. 
}
\end{deluxetable*}

The spectral fitting indicates that some ionization scenarios can be ruled out by the data. 
To quantify how well these scenarios fit the velocity components, we record the difference 
($\delta \chi^2$) of their best-fit $\chi^2$ from those of fitting these lines individually.  
According to an F-test, values $\delta \chi^2<0.1$, 1.0, and 4.0 indicate that a joint fit is 
identical to an individual-line fit at $>94$\% (99\%), 55\% (75\%), and 9\% (19\%) 
confidence levels for Case 3 (Cases 1, 2, 4, 5), respectively.  In this work, a joint fit is
considered acceptable if it is identical to the individual-line fit at $>10\%$ confidence level.  
None of the examined ionization scenarios is acceptable for the blue component, although 
photoionization seems to play an important role in the absorbing gas. The non-fruitful 
exercise for the blue component suggests that the low and high ions are not produced in 
the same absorber in the same ionization conditions (see discussion below). As an 
alternative, non-equilibrium cooling plus photoionization could reconcile all the 
measurements (J. Li 2010, private communication).
Unfortunately, this scenario was not calculated in \citet{gnat07}. 
For the red component, PE and hybrid ionization cases (Cases 2 and 3) are acceptable 
at $>$75\% confidence levels (Table~\ref{tab:results}), the non-equilibrium cooling scenarios 
(Cases 4 and 5) are less acceptable, and the CIE scenario (Case 1) can be ruled out. 

Besides the ion column density ratios, differences in thermal broadening of 
absorption lines of different elements also constrain the gas temperature. 
We assume that the line broadening consists of both thermal and non-thermal 
(turbulent) broadening,  related by $b_D^2 = (b_{\rm ther}^2 + b_{\rm turb}^2)$.  
Taking $kT=1/2mb_{\rm ther}^2$, 
we can use the measured $b_D$ in different elements to determine $T$, 
$b_{\rm ther}$, and $b_{\rm turb}$.  For the red component, measurement
of \hi\ and \ciii\  (Fig.~\ref{fig:line_velocity}) yields $4.9<\log T<5.1$ and 
$0~{\rm km~s^{-1}} < b_{\rm turb}<16~{\rm km~s^{-1}}$, while the combination 
of \hi\ and \ovi\ yields $4.8<\log T<5.1$ and 
$11~{\rm km~s^{-1}}<b_{\rm turb}<22~{\rm km~s^{-1}}$. These constraints
(particularly on temperatures) are consistent with those obtained from
the column density ratios (Table~\ref{tab:results}).

\begin{figure}
  \plotone{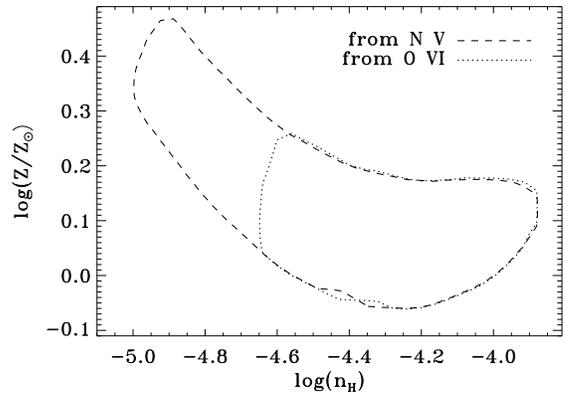}
  \caption{The $1\sigma$ co-varying contours of $\log~(Z/Z_\odot)$ vs.\  $\log~n_H$ 
  for photoionization model, obtained by jointly fitting the blue \hi, \ciii,
  and \siiii\ lines and using the observed amount of \ovi\ (dotted) 
  and \nv\ (dashed) as upper limits.
}
  \label{fig:cont}
\end{figure}

In our analysis, we assumed that both low and high ions were produced in a
co-spatial absorber with the same ionization conditions.  This assumption is often 
justified by the similarity in their absorption-line kinematics 
(Figure~\ref{fig:line_velocity}).   However, it has also been suggested 
(DS08; \citealt{nar10}) that these ions may be multi-phase (non co-spatial), 
with the low ions (e.g., \hi,  \ciii, \siiii) arising from cool-phase photoionized gas 
and the high ions (e.g., \ovi\ and \nv) arising primarily from collisional ionization.
Such scenarios could occur, for instance, in IGM filaments with photoionized cores 
surrounded by infalling shock-heated gas \citep{dan08}.  To examine this possibility, 
we use the PE model (Case 2) to 
jointly fit the blue \hi, \ciii, and \siiii\ lines where all our models failed in joint 
analyses of high and low ion lines (Table~\ref{tab:results}).   Because we have
only two detections (\hi\ and \ciii) and one upper limit (\siiii) and there are three 
free parameters ($n_H$, $Z$, and $N_H$; Table~\ref{tab:results}), we cannot 
obtain meaningful constraints on these parameters. Requiring 
the amount of \ovi\ and \nv\ produced in the PE ionization to be less than those 
observed, $\log~N_{\rm OVI} < 13.52$ and $\log~N_{\rm NV} < 13.30$
 (Fig.~\ref{fig:line_velocity}), we obtain the $1\sigma$ co-varying contour of 
 $\log~(Z/Z_\odot)$ vs.\  $\log~n_H$ (Fig.~\ref{fig:cont}). The overlapped region in
Figure~\ref{fig:cont} indicates the constrained ranges of $\log~n_H = (-4.65, -3.87)$ 
and $\log~(Z/Z_\odot) = (-0.65, 0.25)$ for the cooler PE ionization gas. While the
amount of \ovi\ and \nv\ contained in the PE ionized gas is still poorly constrained, 
it is hard to use the remaining amount to further constrain the properties of 
the hotter collisional gas. Once multiple ionization states of low and high ions are 
available, we can use this method to constrain properties of different phases and 
probe the interactions between them.

All acceptable scenarios require the metallicities of the absorbing gas to
be $\gsim40$\% of the solar value (Table~\ref{tab:results}). Such high
metallicities suggest that we are sampling enriched
intragroup gas or a circumgalactic medium with a small impact distance to a
galaxy (e.g., \citealt{jen05, ara06, pra07}). A deep galaxy survey along the 
1ES~1553+113 sight line (B.\ Keeney \etal\ 2011, in preparation) will 
examine these possibilities and suggest a physical origin for these absorbers.
As the metallicity increases, the cooling times will become shorter compared to 
the recombination time.  The absorbing gas could then become an ``over-cooled'' 
plasma in which the high ions (\civ, \nv, \ovi) are no longer good indicators of the 
gas temperature, especially in cooling models \citep{gnat07}. In this case, 
temperature diagnostics can still be obtained from the differential temperature 
dependence between low ions (e.g., \HI\ and \ciii) and high ions, if they are co-spatial,
or from differential thermal broadening of different elements.  

We assumed relative solar metal abundances of \citet{asp09} in our spectral analysis. 
This assumption can be relaxed when absorption
lines from different ionization states of the same elements are available to constrain 
the physical parameters of the absorber. Spectroscopy of multiple absorption lines 
from different elements will then yield their relative abundances.

The joint analyses presented in this work also enable us to directly estimate
the total baryon content of  \lya\ and metal-line absorbers and the uncertainty 
in those measurements. The primary systematic uncertainties arise from modeling 
the metallicity and ionization fractions.  Since our model uses results of simulations 
and calculations, the metallicity is a known parameter and the ionization corrections 
can easily be calculated.  Once the physical parameters are constrained from 
spectral fitting, we can infer the column density of total hydrogen. 
For example, the best acceptable photoionization and hybrid ionization
scenarios for the red component indicate that the absorber contains a total
$N_{\rm H}=(0.7-1.2)\times10^{18}~{\rm cm^{-2}}$ (Table~\ref{tab:results}).
We are applying this joint analysis technique to systematically study intergalactic 
absorption lines, constrain the physical properties of the absorbers, and estimate 
the total baryons contained within the absorbers. 

In summary, we have presented a physically-based absorption line model for
the spectroscopic study of the IGM.   In this model, we can examine the photoionization
equilibrium, collisional ionization equilibrium, photoionization plus collisional ionization, 
and post-shock gas undergoing non-equilibrium isobaric and isochoric cooling.
For the WHIM, in which $40-50\%$ of the baryons at the current epoch are believed to 
reside, non-equilibrium cooling plus photoionization should be a more proper scenario; 
however, such calculations are not yet available in the literature. For the purpose of 
demonstration, we only simulated a 100-kpc thick (e.g., \citealt{pen02}) intervening cloud
for the absorbing gas along the sight line to 1ES~1553+113.  Because our model directly 
uses the results from calculations and simulations, any progress in future in modeling the 
WHIM and in simulating the 10--100 kpc scales of the circumgalactic medium can easily be
adopted.

\acknowledgements

The authors have benefited from the discussions with Blair Savage, 
Anand Narayanan, and Daniel Wang.  We also thank the referee for
helpful comments and questions.  
This work at the University of Colorado was partly supported by NASA
grant NNX08AC14G for data analysis and scientific discoveries related to the
Cosmic Origins Spectrograph on the Hubble Space Telescope, and by
NNX07AG77G for theoretical work (JMS).  YY also appreciates financial
support by NASA through ADP grant NNX10AE86G.

\end{document}